# 5-Minute Magnetic Particle Spectroscopy (MPS)-based Bioassay: Towards Ultra-fast, Convenient, and Large-scale Disease Diagnosis

Kai Wu[a,†,*], Vinit Kumar Chugh[a,†], Venkatramana D. Krishna[b,†], Yongqiang Andrew Wang[c], Timothy D. Gordon[d], Maxim C-J Cheeran[b,*], and Jian-Ping Wang[a,*]

[a]Department of Electrical and Computer Engineering, University of Minnesota, Minneapolis, MN 55455, United States

[b]Department of Veterinary Population Medicine, University of Minnesota, St. Paul, MN 55108, United States

[c]Ocean Nano Tech LLC, San Diego, CA 92126, United States

[d]Aerosol Devices Inc., Fort Collins, CO 80524, United States

**ABSTRACT:**

In this work, we report a 5-minute magnetic particle spectroscopy (MPS)-based bioassay strategy. In our approach, surface functionalized magnetic nanoparticles (MNPs) are incubated with target analytes at 37 °C with agitation for 3 minutes and the MPS reading is then taken at the 5[th] minute. We have proved the feasibility of 5-minute ultra-fast detection of SARS-CoV-2 spike protein with a detection limit below 5 nM (0.2 pmole). Our proposed 5-minute bioassay strategy may be applied to reduce assay time for other liquid phase, volumetric biosensors such as nuclear magnetic resonance (NMR), quantum dots (QD), fluorescent biosensors, etc.

KEYWORDS: *magnetic particle spectroscopy, magnetic nanoparticle, SARS-CoV-2, bioassay, spike protein, volumetric biosensor*

**MAIN BODY:**

With the ongoing coronavirus disease 2019 (COVID-19) pandemic, there is an urgent need for rapid, convenient, and widely deployable diagnosis tools for the surveillance of severe acute respiratory syndrome-coronavirus-2 (SARS-CoV-2), in a joint effort to mitigate its spread within and across communities. A rapid and sensitive method for early detection of SARS-CoV-2 is critical for controlling the spread of the COVID-19 pandemic by proper containment procedures as well as for reducing morbidity and mortality by facilitating early treatment. Currently, the fastest on-site detection techniques such as lateral flow assays and ID Now can detect COVID-19 in 15 – 30 min. For most on-site diagnosis platforms, the bioassay time varies from 1 hr to 12 hr. As a result, the long turnaround time has severely hindered the COVID-19 surveillance and impeded pandemic control measures.

Magnetic particle spectroscopy (MPS) is an emerging bioassay platform that has been used extensively in the areas of oncology, food safety, bacteria and virus detection. Although it has been reported that portable MPS devices with low assay costs and easy-to-use features can be used for potential field testing, these MPS systems share the same drawback with most point-of-care (POC) diagnosis techniques that is the long bioassay time.

MPS was originally derived from magnetic particle imaging (MPI) in 2006, and has now developed into a mature bioassay technique.[1–4] To date, there are two types of MPS bioassay platforms reported, namely, the surface- and volumetric-based MPS bioassays. Both platforms rely on monitoring the dynamic magnetic responses of magnetic nanoparticles (MNPs), albeit, using different mechanisms. Surface-based MPS bioassay platforms are usually combined with lateral flow strips or non-magnetic porous filters that are surface functionalized to specifically capture target biochemical analytes and MNPs. This surface-based MPS bioassay strategy has been reported for the detection of SARS-CoV-2[5], plant viruses[6], toxins[7–9], and drugs[10]. On the other hand, the volumetric-based MPS bioassay quantifies biochemical analytes through the change of dynamic magnetic responses of freely rotating MNPs before and after the specific binding events. By using specially designed, surface functionalized MNPs, the presence of target biochemical analytes causes different degrees of MNP clustering (see Scheme 1), which impedes the Brownian relaxation of MNPs under an AC magnetic field. Thus, weaker dynamic magnetic responses and lower harmonic amplitudes (i.e., MPS spectra) are observed.[1] This volumetric-based MPS bioassay strategy has been reported for the detection of SARS-CoV-2[11,12], H1N1 virus[13], thrombin and DNA aptamers[14], hormones and cytokines[15].

Compared to surface-based MPS bioassay, this homogeneous and volumetric MPS bioassay strategy can be easily adapted into a one-step, wash-free testing kit for on-site applications, due to its ease of use. The end users simply mix the surface functionalized MNPs with the liquid sample and take MPS readings. However, the bioassay step usually takes 1 hr to 12 hr until the specific binding stabilizes at equilibrium (see step (ii) in Scheme 1). This delay is a major obstacle to the transfer of volumetric MPS bioassays from lab to field testing.

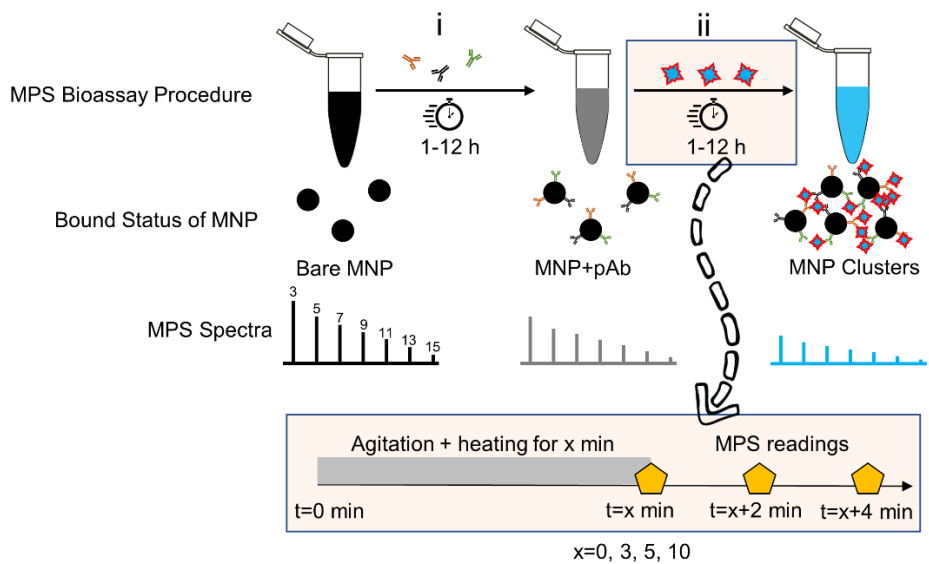

Scheme 1. Incubation conditions applied in order to reduce the MPS bioassay time. (i) MNP surface functionalization with polyclonal antibodies. (ii) Surface functionalized MNPs incubating with target analytes.

Herein, we explored the possibility of reducing the bioassay time by heating and agitating samples during the bioassay step (i.e., step (ii) in Scheme 1). The thermal energy and vibrational kinetic energy (caused by agitation) could increase the frequency of successful collisions between capture probes (i.e., pAb in this work) and target analytes (i.e., SARS-CoV-2 spike protein in this work), thus, allowing for faster specific binding and shorter diagnosis turnaround time. As shown in Table 1, we varied three factors in the incubation condition experiment: time, temperature, and agitation. The incubation time is set at 0 min (control group, no actions taken), 3 min, 5 min, and 10 min. Temperature is set at 25 °C (room temperature, no heating), 32 °C, 37 °C, and 42 °C, respectively, in an incubator. Agitation is applied by placing the tubes in a Vortex Genie 2 mixer (Fisher Scientific Model G-560) set at shaker speed 2. Three independent bioassays are carried out under each incubation condition. A total of 22 experimental groups are designed, where 66 samples are prepared for MPS testing.

The IPG30 MNPs (30 nm iron oxide nanoparticles coated with protein G, 34 nM) are firstly surface functionalized with anti-SARS-CoV-2 spike polyclonal antibodies (rabbit pAb, Cat: 40592-T62, Sino Biological Inc.). As shown in step (i) from Scheme 1. The MNP to pAb ratio is precisely controlled at 1:3, where theoretically each MNP is functionalized with three pAb. This ratio was optimized based on our previous work.[11] Then, 40 μL of pAb functionalized MNP complexes (denoted as 'MNP+pAb' in this work) is mixed with 40 μL, 10 nM SARS-CoV-2 spike protein (Cat: 40592-V08H, Sino Biological Inc.) and incubated under different conditions for x minutes (x=0, 3, 5, and 10). Subsequently, the sample is transferred to the MPS platform at an ambient temperature of 10 °C, three consecutive MPS readings are taken at t=x min, x+2 min, and x+4 min, respectively. The MPS platform consisted of a benchtop system utilizing a pair of magnetic field generation coils, one pick-up coil, data acquisition card by NI, and LabVIEW setup as has been detailed in our previous work.[16] There are two purposes of taking MPS readings in an ambient temperature of 10 °C: 1) to stop the antibody-antigen binding events after x minutes of incubation; 2) the MPS signal is also affected by temperature[17], thus, after incubating under different temperatures, samples should be brought back to the same ambient temperature for MPS readings.

Table 1. Incubation condition experiment design.

| Temperature | Without Agitation | | | With Agitation | | |
|---|---|---|---|---|---|---|
| 25 °C (no heating) | 0 min, control group, no actions taken. (Group 1) | | | 3 min (Group 2) | 5 min (Group 3) | 10 min (Group 4) |
| 32 °C (heating) | 3 min (Group 5) | 5 min (Group 6) | 10 min (Group 7) | 3 min (Group 8) | 5 min (Group 9) | 10 min (Group 10) |

| 37 °C (heating) | 3 min (Group 11) | 5 min (Group 12) | 10 min (Group 13) | 3 min (Group 14) | 5 min (Group 15) | 10 min (Group 16) |
|---|---|---|---|---|---|---|
| 42 °C (heating) | 3 min (Group 17) | 5 min (Group 18) | 10 min (Group 19) | 3 min (Group 20) | 5 min (Group 21) | 10 min (Group 22) |

Figure 1 summarizes the three consecutive MPS readings from samples subjected to different incubation conditions, categorized by the incubation temperatures. Only the 3$^{rd}$ harmonic amplitudes from MPS readings are plotted and compared, as higher harmonics such as the 5$^{th}$, 7$^{th}$, 9$^{th}$, etc., show similar trends. For the control group 1 (solid lines in Figure 1(a)), where no actions are taken during the incubation step, the 3$^{rd}$ harmonic amplitude drops slowly over the 14 min MPS reading window. Since each target SARS-CoV-2 spike protein molecule hosts multiple distinct epitopes that provide specific binding sites for pAb. The presence of target analytes (spike protein) causes the cross-linking of MNPs and hinders the Brownian motion of MNPs as well as weakens the dynamic magnetic response (see Scheme 1). Thus, this drop in real-time harmonic amplitude indicates that antibody-antigen specific binding is taking place, but at a slow rate.

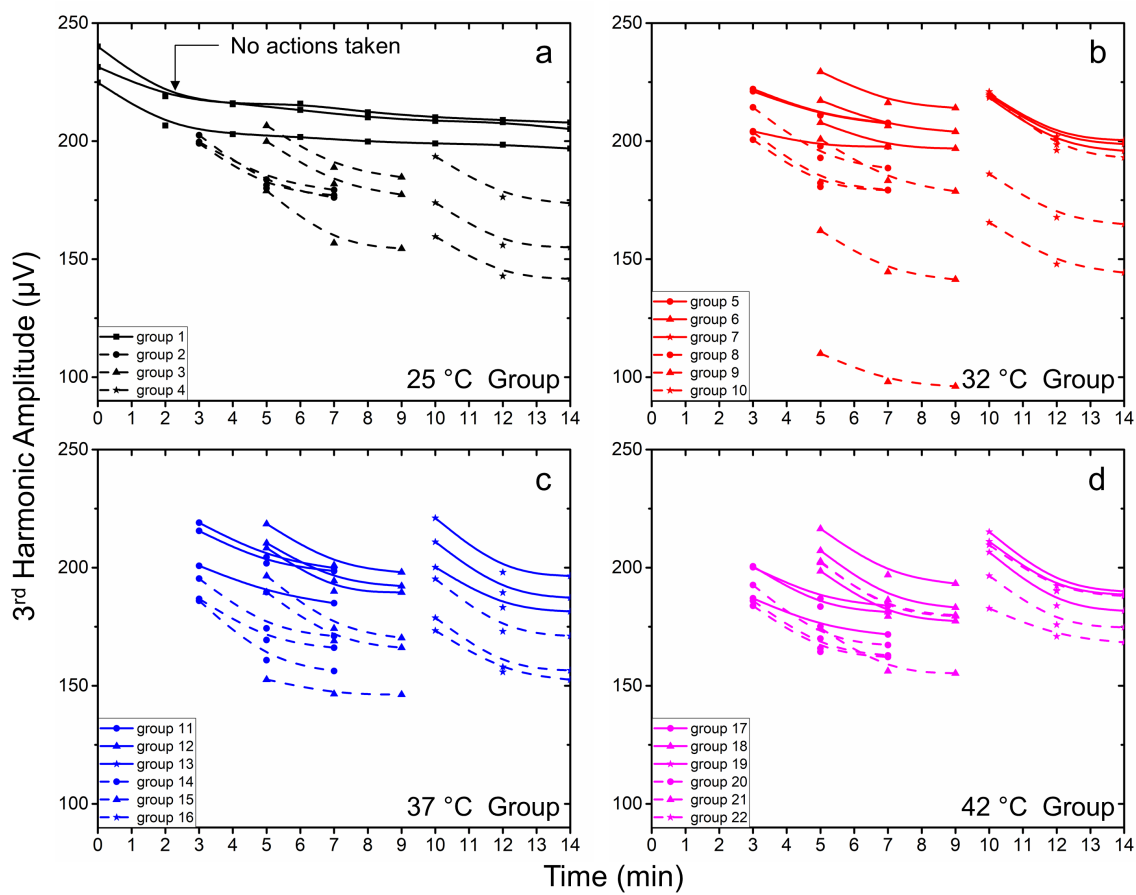

Figure 1. Summary of MPS readings recorded from samples that have undergone different incubation conditions, categorized by incubation temperatures: (a) 25 °C, (b) 32 °C, (c) 37 °C, (d) 42 °C. The first data point of each curve indicates the incubation time. Solid and dashed lines indicate without and with agitation, respectively. The bottom outlier in Group 9 in (b) is caused by air bubbles introduced into the vial during incubation step. This outlier is removed before further data analysis.

Since the lower harmonic amplitudes indicate higher degrees of MNP clustering with more binding events. The dashed lines (with agitation) show lower amplitudes than the solid lines for all temperature groups in Figure 1. Thus, it is concluded that for the same incubation time and temperature, agitation can effectively accelerate antibody-antigen binding. Since all the samples are tested at an ambient temperature of 10 °C, the sudden temperature drop causes a lower harmonic amplitude at t=x+2 min and it becomes stable at t=x+4 min (see Figure 1). Thus, in order to compare the heating effect in the incubation step, the first MPS readings (taken at t=x min) should not be used. Herein, we systematically compared the MPS harmonic signals from all experimental groups in Figure 2. The 3$^{rd}$ harmonic amplitudes are extracted from the second MPS readings (i.e., taken at t=x+2 min) and averaged over three independent bioassays. For all the experimental groups, the harmonic amplitudes are lower than the harmonic amplitude of control group 1 (where no actions are taken during the incubation step, marked as a horizontal line in Figure 2). Without agitation, all experimental samples under heating conditions (i.e., at 32 °C, 37 °C and 42 °C) show lower harmonic amplitudes than control group 1. In addition, for the same incubation time, a higher incubation temperature favors faster antibody-antigen binding, so lower harmonic amplitudes are observed. A longer incubation time favors more antibody-antigen binding events. When agitation is applied, heating can still accelerate antibody-antigen binding. However, if the incubation time is long (such as 5 min or 10 min), the effect of heating becomes less important. For example, by incubating for 3 min with agitation, a higher incubation temperature favors faster binding (as observed by the lower harmonic amplitdues). However, by incubating for 5 or 10 min with agitation, the harmonic amplitude of the heated sample is not significantly different from that of the unheated samples (i.e., at 25 °C).

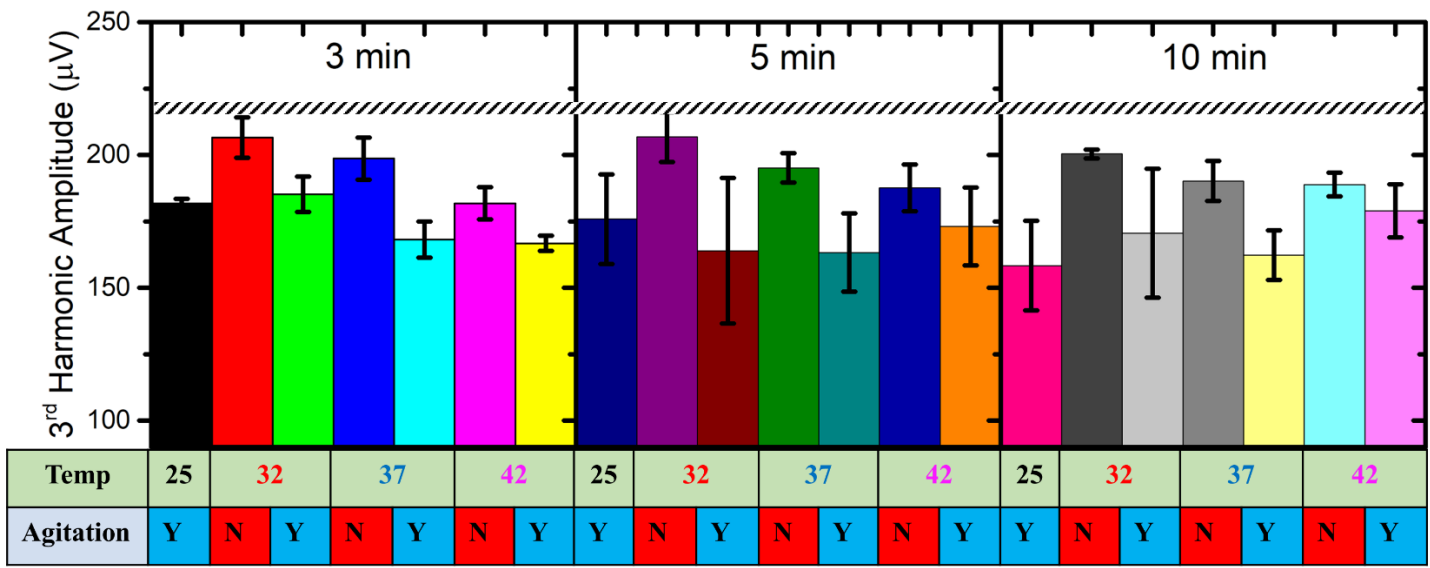

Figure 2. Histograms of the 3rd harmonics recorded at t=x+2 min. The horizontal line at y=220 µV represents the averaged 3rd harmonic amplitude collected from control group 1 where no actions are taken during the incubation step. Signals are averaged over three independent bioassays. Error bars represent standard errors. "N" and "Y" indicate without and with agitation.

From the above discussion. It is concluded that the best incubation conditions to effectively accelerate the binding process is to incubate the sample at 37 °C with agitation for 3 min. Note that, reducing incubation time is our priority; thus, although incubating at 32 °C with agitation for 5 min, or at 37 °C with agitation for 5 min, or at 25 °C with agitation for 10 min shows similar results, they are less favorable than incubation at 37 °C with agitation for 3 min.

Based on the above incubation experiment, we propose an ultra-fast MPS bioassay strategy: (1) incubate samples at 37 °C with agitation for 3 min, (2) transfer the mixture to an ambient temperature of 10 °C and (3) collect the second MPS data point (at t=x+3=5 min). As a proof of concept, we further tested this 5-minute MPS bioassay strategy on different concentrations of SARS-CoV-2 spike protein, from 1000 nM to 0.5 nM. This experiment was conducted on a two-stage lock-in MPS system with an additional voltage gain of around 32 dB for improved bioassay sensitivity. The concentration-response curve is plotted in Figure 3. Five independent bioassays are carried out at each concentration. The 3rd harmonic amplitude saturates at 500 – 1000 nM (upper concentration limit) and 0.5 – 1 nM (lower concentration limit), with a nearly linear response curve between these two limits. As schematically shown in Figure 3, with higher concentrations of SARS-CoV-2 spike protein added, the degree of MNP clustering increases, and the dynamic magnetic response of MNPs becomes weaker thus, lower harmonic amplitudes are observed.

The averaged 3rd harmonic signals from active experimental samples range from 4500 µV to 6000 µV, for samples with SARS-CoV-2 spike protein concentrations varied from 1000 nM to 0.5 nM. For comparison, the 3rd harmonic amplitudes of bare MNPs (IPG30 without pAb functionalization) and pAb functionalized MNPs (denoted as 'MNP+pAb') are 9600 µV and 6000 µV, respectively. Since the pAb conjugated on MNPs impedes the Brownian relaxations, weaker MPS signals are expected from 'MNP+pAb' samples. The detection limit of this 5-minute MPS bioassay for SARS-CoV-2 spike protein is somewhere between 1 nM and 5 nM.

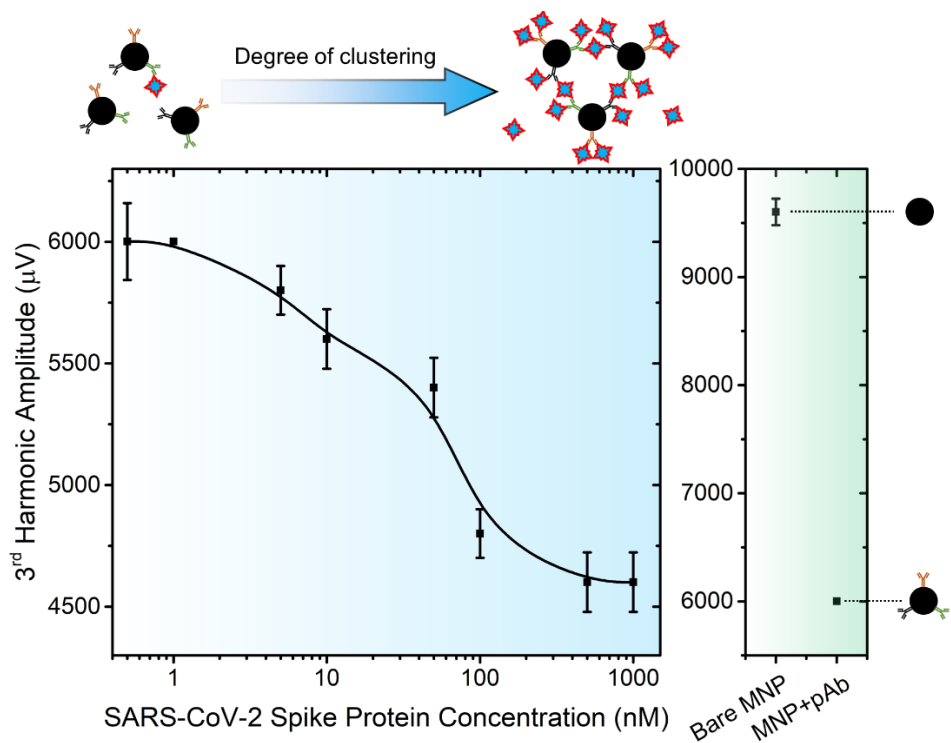

Figure 3. The concentration-response curve of SARS-CoV-2 spike protein tested by a 5-minute MPS bioassay strategy developed in this work. Five independent bioassays are carried out at each concentration. Error bars represent standard errors.

To sum up, based on this work, we propose the application of higher temperatures (37 °C) and agitation conditions during the MPS bioassay incubation step to accelerate the antibody-antigen specific binding. The thermal energy (heating) and vibrational kinetic energy (agitation) increase the frequency of successful collisions between SARS-CoV-2 spike pAbs (from MNP surface) and the spike protein molecules, allowing for faster establishment of specific binding equilibrium and shorter diagnosis turnaround time. Our results show that the 5-minute volumetric MPS bioassay strategy described here could be an effective way to cut the current COVID-19 diagnosis time from 1 hr to 5 minutes. This quick turnaround in diagnosis will greatly advance surveillance and control strategies for diseases especially for future pandemics. Although this proof-of-concept was demonstrated on the volumetric MPS bioassay platform, it can also be applied to other volumetric biosensors such as nuclear magnetic resonance (NMR) biosensor[18,19], ferromagnetic resonance (FMR) biosensor[20], some types of fluorescent biosensors[21], gold nanoparticle-based colorimetric assays[22], etc.

**AUTHOR INFORMATION**
**Corresponding Authors**
*E-mail: wuxx0803@umn.edu (K. W.)
*E-mail: cheeran@umn.edu (M. C-J. C)

*E-mail: jpwang@umn.edu (J.-P. W.)

**ORCID**

Kai Wu: 0000-0002-9444-6112

Vinit Kumar Chugh: 0000-0001-7818-7811

Venkatramana D. Krishna: 0000-0002-1980-5525

Yongqiang Andrew Wang: 0000-0003-2132-2490

Timothy D. Gordon: 0000-0002-5128-9532

Maxim C-J Cheeran: 0000-0002-5331-4746

Jian-Ping Wang: 0000-0003-2815-6624

**Author Contributions**

[†]K.W., V.K.C., and V.D.K. have contributed equally to this work.

**Notes**

The authors declare no conflict of interest.

**ACKNOWLEDGMENTS**

This study was financially supported by the Institute of Engineering in Medicine, the Robert F. Hartmann Endowed Chair professorship, the University of Minnesota Medical School, and the University of Minnesota Physicians and Fairview Health Services through COVID-19 Rapid Response Grant. This study was also financially supported by the U.S. Department of Agriculture - National Institute of Food and Agriculture (NIFA) under Award Number 2020-67021-31956. Research reported in this publication was supported by the National Institute Of Dental & Craniofacial Research of the National Institutes of Health under Award Number R42DE030832. The content is solely the responsibility of the authors and does not necessarily represent the official views of the National Institutes of Health. Portions of this work were conducted in the Minnesota Nano Center, which is supported by the National Science Foundation through the National Nano Coordinated Infrastructure Network (NNCI) under Award Number ECCS-1542202.

**TOC:**

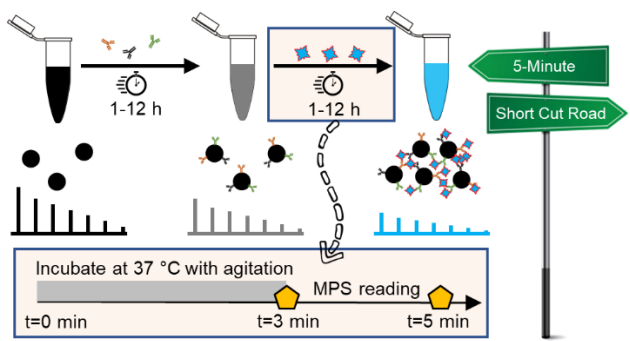